\documentclass[12pt]{article}
\usepackage[english]{babel}
\usepackage[utf8]{inputenc}
\usepackage{johd}
\usepackage{ragged2e}
\usepackage{amsmath}
\usepackage[table]{xcolor}     % colour support
\definecolor{darkgreen}{RGB}{0,100,0}
\definecolor{golden}{RGB}{218,165,32}   % deep golden / dark‑yellow
\usepackage{threeparttable} 
\usepackage{rotating}
\usepackage{parskip}
\usepackage{array}
\usepackage{graphicx}
\usepackage{times}
\usepackage{caption}
\usepackage{multirow}
\usepackage{booktabs}
\usepackage{tikz}
\usetikzlibrary{shapes.geometric, arrows.meta, positioning}
\usepackage{amssymb}
\usepackage{subfig}
\usepackage{natbib}
\usepackage[export]{adjustbox}
\usepackage{wrapfig}
\usepackage{titlesec}
\titleformat{\section}
  {\normalfont\Large\bfseries\centering}{\thesection}{1em}{}
\titleformat{\subsection}
  {\normalfont\large\bfseries\centering}{\thesubsection}{1em}{}
\usetikzlibrary{calc}
\tikzset{
  block/.style = {rectangle, rounded corners, draw, minimum width=3cm, minimum height=1cm, align=center},
  arrow/.style = {-{Stealth}, thick}
}
  
\title{\textbf{Behavioral Probability Weighting and Portfolio Optimization under Semi-Heavy Tails}}

\author{Ayush Jha$^{1}$$^{*}$ \;\; Abootaleb Shirvani$^{2}$ \;\; Ali Jaffri$^{3}$  \\ \\ Svetlozar T. Rachev$^{4}$  \;\;  Frank J. Fabozzi$^{5}$ \\ \\
        \small $^{1}$Department of Economics, Texas Tech University \\
        \small $^{2}$Department of Mathematical Sciences, Kean University \\
        \small $^{3}$College of Business, North Dakota State University \\
        \small $^{4}$Department of Mathematics and Statistics, Texas Tech University \\
        \small $^{5}$Carey Business School, Johns Hopkins University \\\\
        \small $^{*}$Corresponding author: Ayush Jha; \tt{ayush.jha@ttu.edu} \\
}
\date{}

\begin{document}
\maketitle
\begin{abstract} 
\noindent This paper develops a unified framework that integrates behavioral distortions into rational portfolio optimization by extracting implied probability weighting functions (PWFs) from optimal portfolios modeled under Gaussian and Normal-Inverse-Gaussian (NIG) return distributions. Using DJIA constituents, we construct mean-CVaR$_{99}$ frontiers, alongwith Sharpe- and CVaR-maximizing portfolios, and estimate PWFs that capture nonlinear beliefs consistent with fear and greed. We show that increasing tail fatness amplifies these distortions and that shifts in the term structure of risk-free rates alter their curvature. The results highlight the importance of jointly modeling return asymmetry and belief distortions in portfolio risk management and capital allocation under extreme-risk environments.
\end{abstract}

\noindent\\

\noindent
\setlength{\parskip}{10pt}
\newpage
\section{Introduction}

Optimal asset allocation hinges on the interplay between rational risk–return trade‐offs and the behavioral impulses of greed and fear that shape investors’ subjective perceptions of uncertainty. In classical mean–variance theory, pioneered by \citet{Markowitz1952} and extended by \citet{Sharpe1964}, investors maximize the Sharpe ratio under objective probability weighting, yielding a unique tangency portfolio along the Capital Market Line. Yet real‐world returns exhibit skewness and leptokurtosis \citep{KrausLitzenberger1976,HarveySiddique2000}, and empirical allocations deviate from CAPM predictions, giving rise to persistent anomalies such as the equity premium puzzle \citep{MehraPrescott1985} and the low‐volatility effect \citep{AngChen2002}. Prospect theory \citep{KahnemanTversky1979} and its cumulative extension \citep{TverskyKahneman1992} offer a behavioral alternative, positing that investors apply inverse‐S probability weighting functions (PWFs) that overweight rare tail events and underweight moderate ones. These distortions, confirmed experimentally by \citet{Prelec1998}, underpin behavioral phenomena such as the disposition effect \citep{ShefrinStatman1985,Odean1998} and lottery‐like demand for high‐skew assets \citep{BaliCakiciWhitelaw2011}, inducing systematic biases in portfolio allocations.

Recent theoretical contributions incorporate such PWFs into optimization models. \citet{BarberisHuang2008} and \citet{BordaloGennaioliShleifer2013} show that rank‐dependent utility generates “pyramidal” portfolios that combine safety and speculation, while \citet{Polkovnichenko2005} attributes severe under‐diversification to probability weighting alone. However, these models often impose restrictive parametric forms, typically Gaussian or exponential, that fail to capture the heavy tails and jump risk documented in empirical return distributions \citep{KellyJiang2014}. Separately, tail‐risk measures like Conditional Value‐at‐Risk (CVaR) have been introduced into the optimization framework \citep{RockafellarUryasev2000}, and studies such as \citet{AngChen2002} show that CVaR-based portfolios tilt toward safer assets compared to their Sharpe‐optimal counterparts. Yet, even these approaches typically assume objective probability weighting, leaving a gap in frameworks that integrate behavioral distortions, through PWFs, into both variance‐ and tail‐risk–based portfolio construction.

In this paper, we bridge rational and behavioral paradigms by extracting implied probability-weighting functions (PWFs) from optimal portfolios estimated under heavy‐tailed return distributions. Using the thirty constituents of the Dow Jones Industrial Average, we construct mean–variance and CVaR$_{99}$ efficient frontiers and extending the analysis to the Sharpe-maximizing and CVaR-maximizing tangent portfolios. Under two comparative distributions, Gaussian and Normal Inverse Gaussian (NIG), we derive the PWFs implied by each portfolio's tangency point used as the subjective view relative to the objective or prior view set as the DJIA index benchmark. This approach quantifies the behavioral distortion embedded in optimal allocations under varying assumptions about return asymmetry and kurtosis.

Our results reveal that heavier-tailed laws produce PWFs with sharper inverse‐S curvature, capturing heightened fear (overweighting of extreme losses) and greed (overweighting of extreme gains). These effects are magnified under CVaR$_{99}$ optimization relative to mean–variance, indicating that tail-risk minimization intensifies behavioral distortions. Additionally, we show that shifts in the term structure of risk-free rates alter the shape of the implied PWF, suggesting that funding conditions interact with behavioral risk perceptions. Together, our framework offers a tractable method for identifying and managing behavioral risk in portfolio choice, grounded in observable return distributions.

\section{Methodology} \label{methods}

\subsection{Data}

Our empirical analysis proceeds in two parts: (i) classical portfolio optimization under the CVaR criteria, and (ii) forward-looking evaluation of Sharpe-maximizing and CVaR$_{99}$-maximizing tangency portfolios relative to a DJIA benchmark. This section outlines the data construction and transformation steps supporting each component, along with the distributional diagnostics used as robustness checks.

\subsubsection{Portfolio Analysis}

We construct an equity portfolio comprising the thirty constituents of the Dow Jones Industrial Average (DJIA), a price-weighted index of U.S. large-cap equities. The dataset includes unadjusted closing prices obtained from Bloomberg as of June 14, 2025, covering the period from November 19, 2020 to May 3, 2025, yielding five years of daily data. These unadjusted series exclude dividend effects and reflect capital appreciation alone, ensuring return calculations isolate price dynamics.

Daily arithmetic returns are computed from raw prices, consistent with discrete-time portfolio analysis and convex optimization requirements. As noted in \citet{Shiryaev1999}, arithmetic returns preserve aggregation properties and allow tractable solutions for CVaR frameworks. This transformation also mitigates nonstationarity in raw price levels and stabilizes risk estimates across assets.

We employ two proxies for the risk-free rate: the three-month and ten-year U.S. Treasury yields, reported as annualized, continuously compounded rates. These are converted to effective daily rates via \( r_{f,3m} = 0.042/252 \) and \( r_{f,10y} = 0.046/252 \), respectively, and used in the construction of capital market lines and tangency portfolios across optimization regimes.

\subsection{Tangent Portfolio Analysis}

Tangent portfolios lie at the core of modern asset allocation, identifying optimal trade-offs between risk and return. Under Gaussian assumptions and the availability of a risk-free asset, the tangency portfolio maximizes the Sharpe ratio along the Capital Market Line (CML). In practice, asset returns display skewness, kurtosis, and discontinuities, limiting the effectiveness of variance-based metrics. Moreover, tail-sensitive investors often favor risk measures such as Conditional Value-at-Risk (CVaR). As such, we estimate tangency portfolios under both variance and CVaR objectives and assess their empirical performance in- and out-of-sample.

\begin{tikzpicture}[node distance=1cm and 2cm]

  % 1. Input
  \node[block] (input) {Input: Return series\\$r_t$ (window $W$)};

  % 2. Estimate
  \node[block, below=of input] (est) {Step 1: Estimate $\widehat\mu$ \& $\widehat\Sigma$\\(Mean and Covariance Matrix)};

  % 3a. Sharpe branch
  \node[block,
        below left=1.5cm and 1.5cm of est] (sharpe)
    {Step 2a: Sharpe‐optimal\\
     $\displaystyle\max_{w\ge0,\;\mathbf1^\top w=1}\frac{\widehat\mu^\top w}{\sqrt{w^\top\widehat\Sigma w}}$};

  % 3b. CVaR branch
  \node[block,
        below right=1.5cm and 1.5cm of est] (cvar)
    {Step 2b: CVaR‐optimal\\
     $\displaystyle\max_{w\ge0,\;\mathbf1^\top w=1}\frac{\widehat\mu^\top w}{\mathrm{CVaR}_{0.99}(w)}$};

  % 4. Rolling
  % place under the midpoint of the two branches
  \coordinate (mid) at ($(sharpe)!0.5!(cvar)$);
  \node[block, below=of mid] (roll) {Rolling 4-year window};

  % 5. Out-of-sample
  \node[block, below=of roll] (oos)
    {Generate 100-day out-of-sample returns\\ vs.\ DJIA benchmark};

  % Arrows
  \draw[arrow] (input) -- (est);
  \draw[arrow] (est) -- (sharpe);
  \draw[arrow] (est) -- (cvar);
  \draw[arrow] (sharpe) -- (roll);
  \draw[arrow] (cvar) -- (roll);
  \draw[arrow] (roll) -- (oos);

\end{tikzpicture}

\subsection{Distributional Fit and Out-of-Sample Performance}

To evaluate the statistical adequacy of competing return distributions, we fit Normal and Normal-Inverse-Gaussian (NIG) laws to the 100-day out-of-sample returns of Sharpe-maximizing, CVaR$_{99}$-maximizing, and DJIA benchmark portfolios. Table~\ref{tab:metrics} reports the AIC, BIC, and Kolmogorov–Smirnov (KS) statistics across specifications. In all cases, the NIG distribution dominates on both likelihood-based and nonparametric metrics, reflecting its superior ability to accommodate skewness and excess kurtosis observed in realized returns.

The difference is particularly notable for the DJIA and CVaR portfolios, where Gaussian fits significantly understate tail risk. In the Sharpe-optimized portfolio, the estimated NIG parameters imply a pronounced tail index (\(\alpha \approx 0.39\)), capturing discrete jump behavior that standard Normal distributions cannot replicate. This divergence has immediate economic significance: optimization under mis-specified distributions leads to incorrect asset weights and underestimation of risk exposures, particularly in turbulent regimes.

\begin{table}[htbp]
\centering
\begin{tabular}{llrrr}
\toprule
\toprule
Portfolio & Distribution     &    AIC     &    BIC     & KS statistic \\
\cmidrule{3-5} \\
\multirow{2}{*}{Sharpe}
 & Normal          &  -553.411  &  -548.201  & 0.10450 \\
 & NIG             &  -590.036  &  -579.615  & 0.04144 \\
\midrule
\multirow{2}{*}{CVaR}
 & Normal          &  -498.331  &  -493.121  & 0.10883 \\
 & NIG             &  -514.163  &  -503.742  & 0.05381 \\
\midrule
\multirow{2}{*}{DJIA}
 & Normal          &  -577.127  &  -571.917  & 0.19146 \\
 & NIG             &  -660.546  &  -650.125  & 0.05142 \\
\bottomrule
\bottomrule
\end{tabular}
\caption{Goodness‐of‐fit metrics by distribution and portfolio}
\label{tab:metrics}
\end{table}

\begin{figure}[htbp]
  \centering
  \begin{minipage}[b]{0.49\textwidth}
    \centering
    \includegraphics[width=\textwidth]{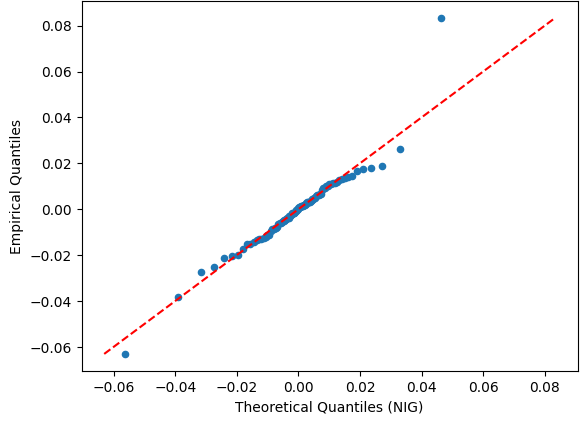}
    \caption{Sharpe-Max vs. NIG}
    \label{fig:sharpe-qq-nig}
  \end{minipage}
  \hfill
  \begin{minipage}[b]{0.49\textwidth}
    \centering
    \includegraphics[width=\textwidth]{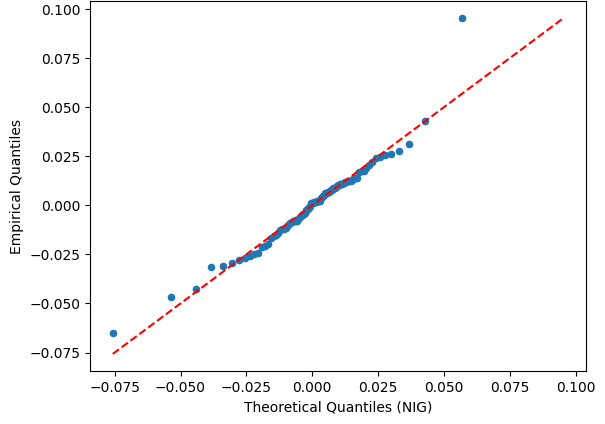}
    \caption{CVaR-Max vs. NIG}
    \label{fig:cvar-qq-nig}
  \end{minipage}
\end{figure}

Out-of-sample diagnostics further reinforce the validity of the NIG model. Figures~\ref{fig:sharpe-qq-nig} and \ref{fig:cvar-qq-nig} display QQ-plots for the Sharpe and CVaR portfolios. In both cases, the NIG distribution aligns closely with empirical quantiles across the entire support, including both tails. This alignment is critical for forward-looking portfolio design, where accurate tail modeling influences capital allocation, drawdown control, and investor utility.

\begin{figure}[htbp]
    \centering
    \includegraphics[width=1\linewidth]{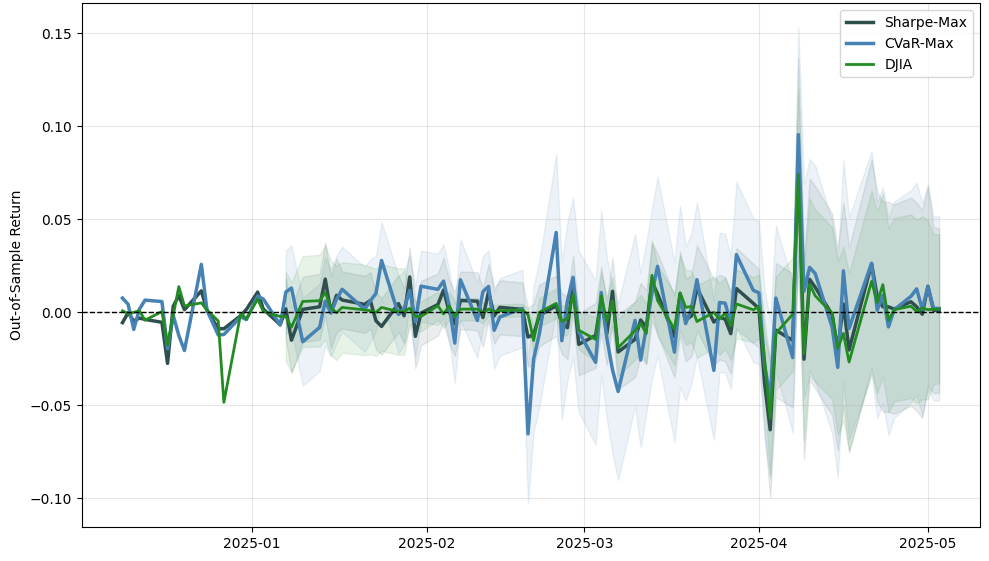}
    \caption{Out-of-Sample Expected Returns of the Three Portfolios}
    \label{fig:oosreturns}
\end{figure}

Figure~\ref{fig:oosreturns} illustrates the return paths of all three portfolios. A noticeable divergence emerges during Q1 2025, where increased downside risk materializes more acutely in the CVaR-maximizing strategy. Following a dovish Fed pivot in April 2025, all portfolios rebound, with the CVaR allocation exhibiting the sharpest upside, consistent with its tail-sensitive optimization structure. These shifts emphasize the importance of modeling return distributions that can dynamically respond to market regime changes.

These results form the basis for incorporating probability-weighting functions (PWFs) under Prospect Theory. By aligning realized returns with their distributional priors and mapping through PWFs, we translate statistical tail asymmetries into behavioral terms. In the next section, we quantify greed and fear via the curvature and slope of implied PWFs derived under Gaussian, NIG, and other semi-heavy tailed laws. The shape of each function, particularly under NIG priors, provides a direct economic measure of how investors overweight or underweight return outcomes, shaping portfolio construction in both rational and behavioral dimensions.

\subsection{Probability‐Weighting Functions under Distributional Distortions}

We quantify belief distortions by constructing probability-weighting functions (PWFs) from the composition of subjective and objective return distributions. Using the DJIA as the rational benchmark, each optimized portfolio induces a subjective distribution $F_s$ over returns. The implied PWF maps objective quantiles $u \in [0,1]$ to distorted probabilities:
\[
  w_s(u)
  = F_s\bigl(F_R^{-1}(u)\bigr)
  = \Phi\!\Bigl(\frac{\mu_R + \sigma_R\,\Phi^{-1}(u)-\mu_s}{\sigma_s}\Bigr)
\]
under distribution-specific assumptions. We extend this to two comparative return distribution-based beliefs, Normal and NIG, to evaluate how tail fatness amplifies behavioral distortions. As kurtosis and skewness rise, the curvature of $w_s(u)$ steepens, reflecting greater overweighting of rare losses (fear) and rare gains (greed).

\vspace{1em}
\begin{center}
\begin{tikzpicture}[node distance=1.2cm, every node/.style={align=center}, scale=0.95, transform shape]
  \tikzset{
    block/.style={
      rectangle, draw=black, rounded corners,
      minimum width=3.8cm, minimum height=1cm,
      fill=orange!40
    },
    arrow/.style={->, thick},
  }

  % Nodes (top to bottom)
  \node[block] (u) {Objective Quantile\\$u \in [0,1]$};
  \node[block, below=of u] (F_R_inv) {Inverse CDF:\\$F_R^{-1}(u)$};
  \node[block, below=of F_R_inv] (F_s) {Subjective CDF:\\$F_s(\cdot)$};
  \node[block, below=of F_s] (wsu) {PWF:\\$w_s(u)$};

  % Arrows
  \draw[arrow] (u) -- (F_R_inv) node[midway, right] {\small DJIA prior};
  \draw[arrow] (F_R_inv) -- (F_s) node[midway, right] {\small Portfolio-specific fit \\ (Normal vs. NIG)};
  \draw[arrow] (F_s) -- (wsu) node[midway, right] {\small Implied distortion};
\end{tikzpicture}
\end{center}
\vspace{1em}

This structure isolates the behavioral component in portfolio allocation, revealing how semi-heavy-tailed beliefs, particularly under NIG priors, endogenously generate inverse-S shaped distortions consistent with Prospect Theory.

\section{Results} \label{results}

\subsection{Portfolio Overweighting vs. Underweighting}

\begin{figure}[htbp]
    \centering
    \includegraphics[width=0.9\linewidth]{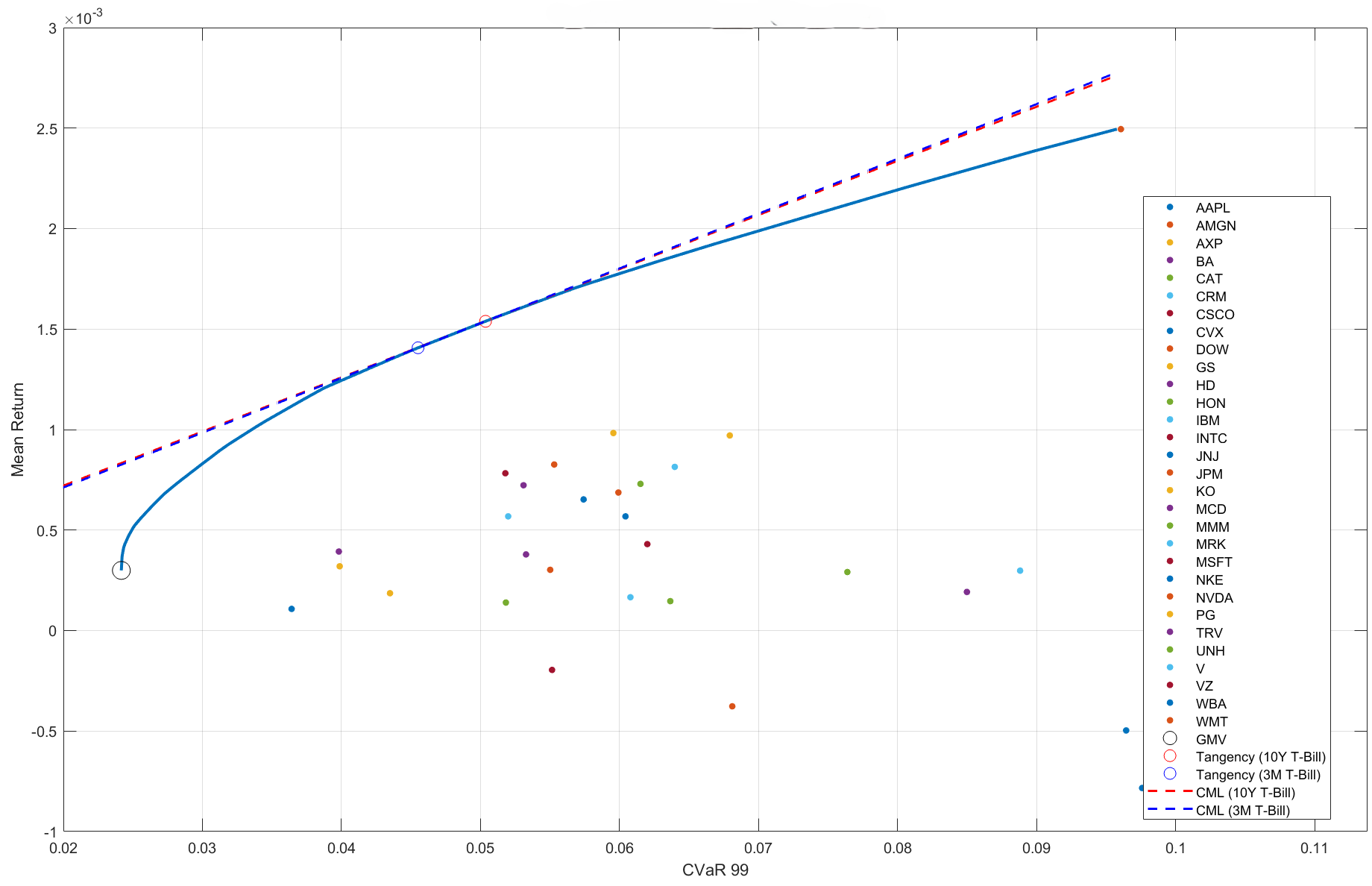}
    \caption{Mean-CVaR$_{99}$ Efficient Frontier}
    \label{fig:MEF_cvar99}
\end{figure}

To explore tail-risk aversion, we construct the CVaR$_{99}$ frontier in Figure~\ref{fig:MEF_cvar99} using convex optimization over long-only portfolios. This frontier maps the lowest possible expected shortfall, defined as the average loss beyond the 99th percentile, for each level of expected return. The global minimum-CVaR portfolio minimizes exposure to extreme left‐tail events, while each tangency portfolio maximizes the expected return per unit of CVaR. Economically, these tangency solutions reveal how optimal portfolio allocations adjust to shifts in the term structure of interest rates.

\begin{figure}[htbp]
  \centering
  % First image
  \begin{minipage}[b]{0.49\textwidth}
    \centering
    \includegraphics[width=\textwidth]{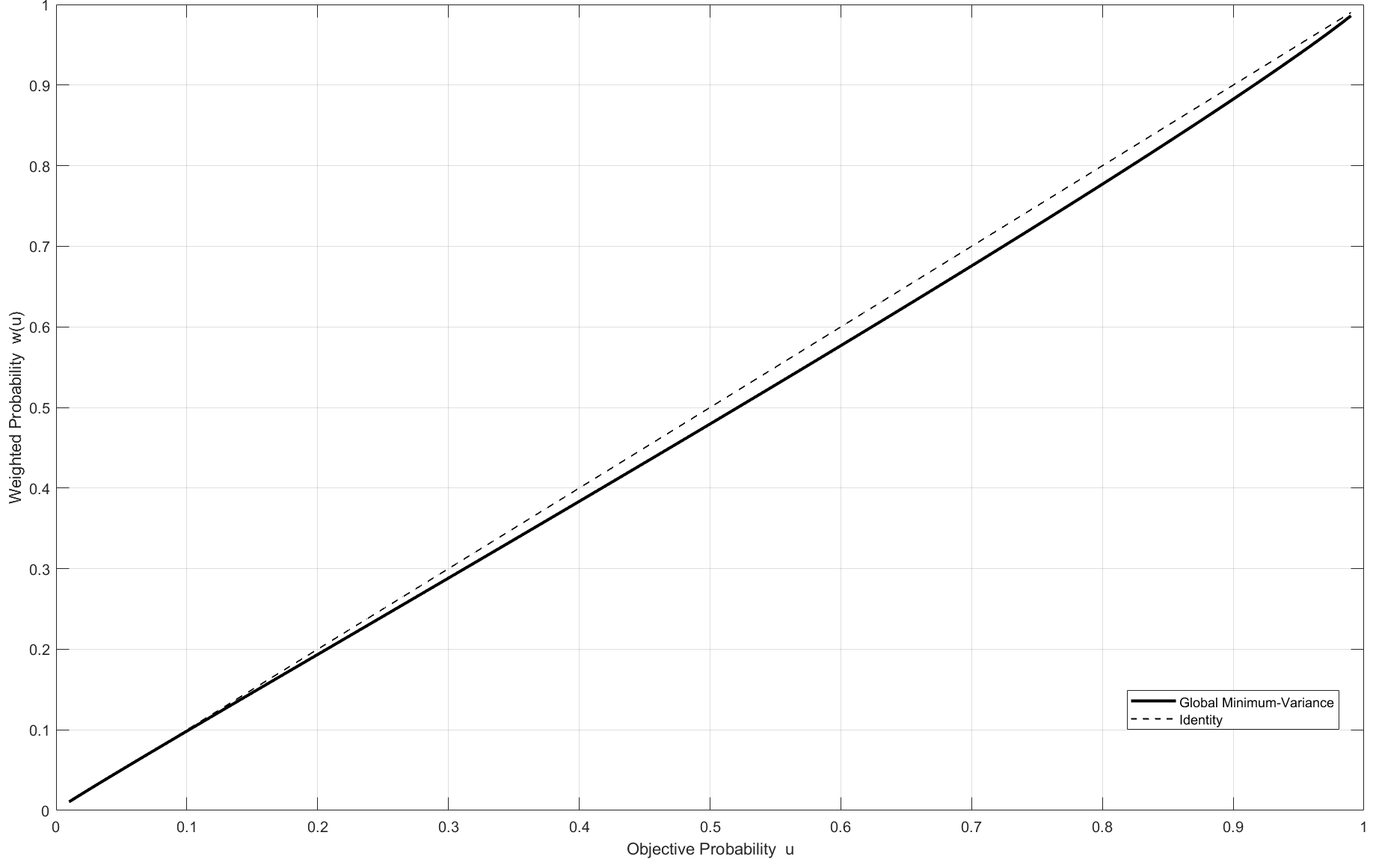}
    \caption{Global Min-CVaR$_{99}$}
    \label{fig:gmv_cvar99}
  \end{minipage}
  \hfill
  % Second image
  \begin{minipage}[b]{0.49\textwidth}
    \centering
    \includegraphics[width=\textwidth]{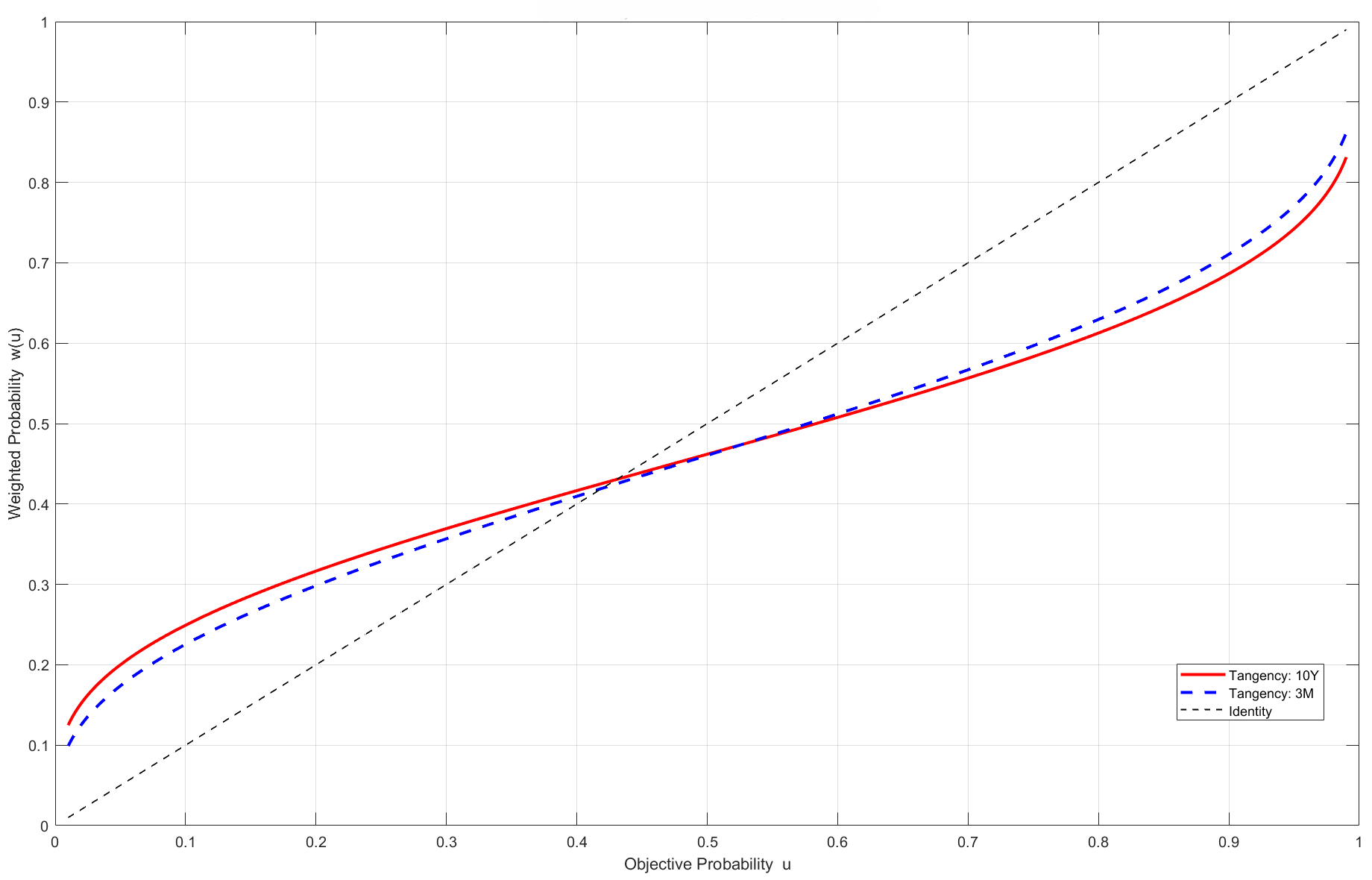}
    \caption{CVaR$_{99}$ Tangent Portfolios}
    \label{fig:tang_cvar99}
  \end{minipage}
\end{figure}

The Gaussian PWF for the minimum‐CVaR portfolio, shown in Figure~\ref{fig:gmv_cvar99}, underweights moderate events and approaches linearity at the tails, reflecting heightened caution toward intermediate risk and near-objective evaluation of extreme outcomes. Tangency portfolios under CVaR optimization retain the inverse‐S distortion, stronger for the ten-year benchmark, demonstrating that tail-risk metrics intensify probability weighting in both directions.

Figure~\ref{fig:tang_cvar99} overlays the Gaussian PWFs for the two tangency portfolios under CVaR99.  Both curves display the characteristic inverse‐S shape: they rise above the identity line at low objective probabilities, signaling overweighting of extreme losses (fear), then bend below it for middle quantiles, before arching back above identity in the upper tail, capturing greed for rare gains.  The ten‐year tangency curve exhibits slightly greater convexity in both tails, indicating stronger distortions of low‐probability events when the long‐term risk‐free rate is higher.  Conversely, the three‐month curve is closer to linear, suggesting that a lower short-term benchmark attenuates both fear and greed. This comparative divergence reveals how changes in the term structure of interest rates can amplify or dampen behavioral biases in tail‐risk perception.

\begin{figure}
    \centering
    \includegraphics[width=0.7\linewidth]{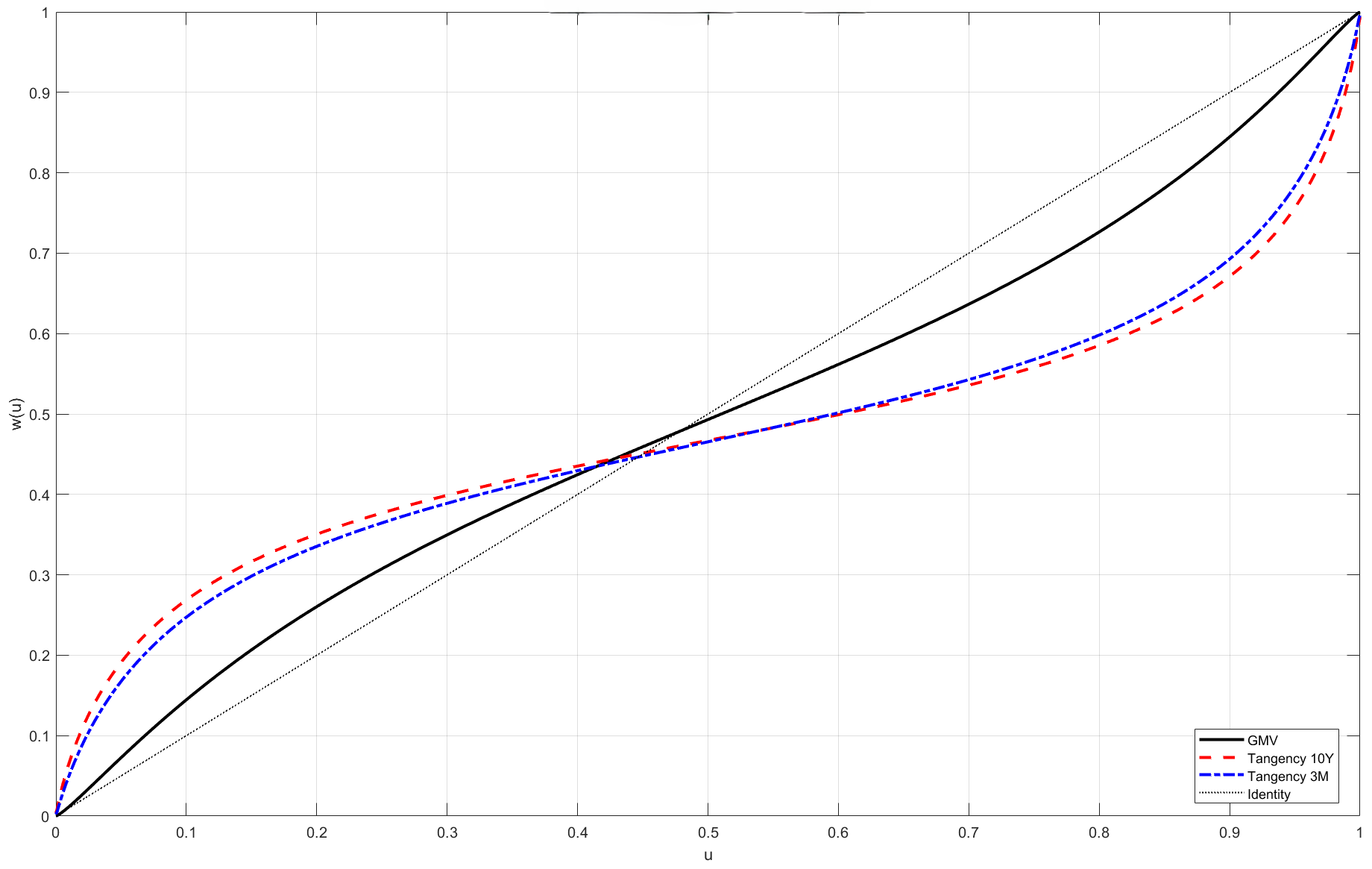}
    \caption{CVaR$_{99}$ Tangency-NIG}
    \label{fig:nig_99}
\end{figure}

In contrast, the NIG-based PWF (Figure~\ref{fig:nig_99}) reflects substantial nonlinearities in belief formation. The heavier tails of the NIG distribution induce steep over-weighting of both extreme losses and extreme gains. amplifying behavioral expressions of fear and greed. This curvature is not imposed externally, but rather emerges endogenously from the interaction between statistical tail risk and the optimization objective. The implication is that under more realistic return models, even rational investors minimizing CVaR behave as if they systematically distort low-probability outcomes, overweighting downside risk far more than a Normal model would predict.

In conclusion, these results integrate the rational logic of CVaR‐based portfolio choice with the behavioral insights of prospect‐theoretic probability weighting. The CVaR99 frontier identifies the portfolios that rationally minimize tail losses, while the Gaussian PWFs introduce systematic distortions, fearful overweighting of moderate losses and greedy overweighting of extreme gains, that reconcile the observed gap between mean-tail optimization and actual investor behavior. By embedding these nonlinear beliefs into the frontier analysis, we bridge normative risk‐management with behavioral distortions, offering a unified framework for understanding how practitioners navigate extreme‐risk environments.

\subsection{Reconciling Behavioral and Rational Finance in Tangent Portfolios}

We examine how probability-weighting functions (PWFs), estimated from forecasted return distributions, vary across portfolio types and statistical priors. Using the Sharpe-maximizing, CVaR$_{99}$-maximizing, and DJIA benchmark portfolios, we compare distortions in subjective probability perception under the same distributions as in the previous section. This framework embeds prospect-theoretic distortions within standard portfolio optimization and quantifies how rationally optimal portfolios may implicitly encode behavioral preferences when return forecasts exhibit skewness or heavy tails.

\begin{figure}[htbp]
  \centering
  \begin{minipage}[b]{0.49\textwidth}
    \centering
    \includegraphics[width=\textwidth]{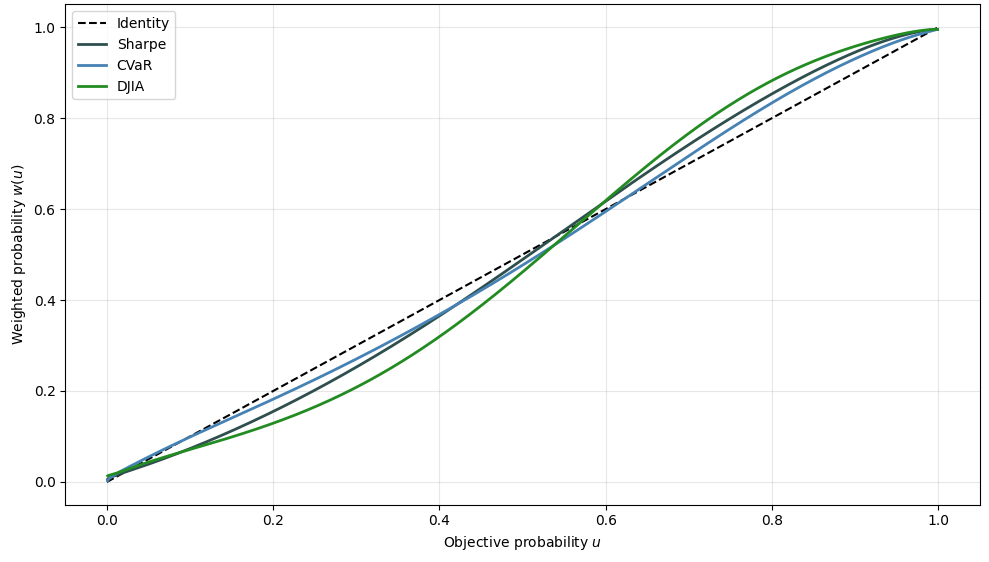}
    \caption{Normal PWF}
    \label{fig:cvar99_dow}
  \end{minipage}
  \hfill
  \begin{minipage}[b]{0.49\textwidth}
    \centering
    \includegraphics[width=\textwidth]{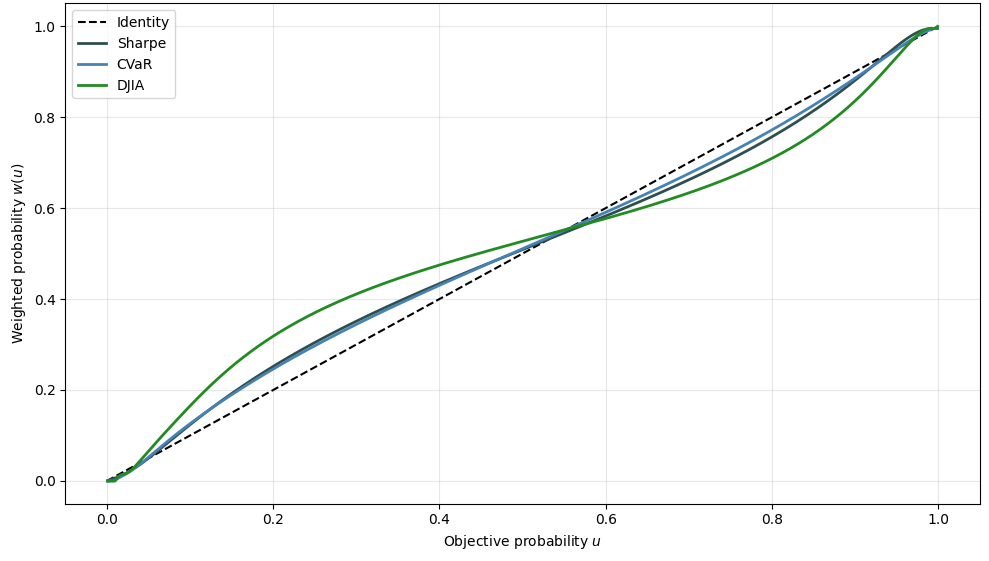}
    \caption{NIG PWF}
    \label{fig:cvar99_cryp}
  \end{minipage}
\end{figure}

Under Gaussian innovations, all portfolios exhibit modest inverse-S weighting, slight fear in the loss tail and modest greed in the gain tail, consistent with limited probability distortion under light-tailed dynamics. In contrast, NIG-based PWFs display steeper curvature: extreme gains are substantially overweighted and losses disproportionately underweighted. This endogenous distortion is strongest for the Sharpe-maximizing portfolio, suggesting that optimal allocations, when paired with semi-heavy-tailed beliefs, amplify behavioral asymmetries beyond those embedded in benchmark indices like the DJIA.

These patterns carry important implications for portfolio risk management and capital planning. Probability distortions induced by return asymmetry reshape investors’ sensitivity to extreme outcomes, thereby altering portfolio responses to downside risk. For example, a steeper PWF curvature implies that, from the investor’s subjective perspective, rare losses are perceived as more likely. This can lead to more conservative capital buffers, heightened Value-at-Risk sensitivity, and reduced tolerance for drawdowns, even when statistical volatility remains unchanged. In terms of liquidity, portfolios optimized under heavy-tailed beliefs may demand greater cash or near-cash reserves to accommodate abrupt rebalancing triggered by tail-risk realizations. These behaviors, while often interpreted as overreactions, may instead reflect rational adjustments to belief-weighted risk exposure, highlighting that effective risk management must jointly account for distributional assumptions and behavioral weighting of tail events.

Our findings emphasize that behavioral features in portfolio choice may be less about psychological deviations from rationality, and more about rational responses to non-Gaussian return environments. Accurate estimation of PWFs therefore requires not only a behavioral lens but also correct distributional calibration. Integrating semi-heavy-tailed priors like NIG provides a coherent bridge between the behavioral weighting of tail events and the econometric structure of returns, yielding a tractable synthesis of behavioral and rational finance.

\section{Conclusion}

This paper presents a unified framework that embeds prospect-theoretic probability weighting into classical portfolio optimization under alternative distributional assumptions. By estimating probability-weighting functions (PWFs) from optimal portfolios constructed under Normal and NIG return innovations, we demonstrate how behavioral distortions, typically treated as exogenous, can arise endogenously from investors' rational responses to asymmetric and heavy-tailed return environments. In particular, we show that optimized portfolios under heavy-tailed priors exhibit significantly steeper inverse-S PWF curvature, reflecting heightened sensitivity to rare loss and gain outcomes. These distortions intensify under tail-risk objectives such as CVaR$_{99}$ and are most pronounced in Sharpe-maximizing portfolios calibrated with semi-heavy-tailed models like NIG. Our results highlight that behavioral features such as fear and greed can be structurally linked to distributional calibration and optimization targets, offering a tractable bridge between rational and behavioral finance. Beyond theory, this has direct implications for how investors assess capital at risk, manage liquidity buffers, and allocate resources under stress scenarios, suggesting that modern risk management must consider not only volatility and expected shortfall, but also the belief-weighted probability structure that governs investor behavior under uncertainty.

\end{document}